# Método basado en espectrogramas de Mel y expresiones regulares para la identificación en tiempo real de vehículos de emergencia

Pacheco, Alberto*; Torres, Raymundo*; Chacón, Raúl*; Robledo, Isidro*
*Tecnológico Nacional de México campus Chihuahua
División de Estudios de Posgrado e Investigación
Av. Tecnológico 2909, Col. 10 de mayo. Chihuahua, Chih.
Tel. (614) 201-2000 ext. 2112
{alberto.pg, m21061139, raul.cb, robledo.rv} @chihuahua.tecnm.mx

**RESUMEN.**
En situaciones de emergencia, el desplazamiento de ambulancias por las vialidades de una urbe puede ser problemático debido al tráfico vehicular. El presente trabajo presenta un método para la detección de sirenas de vehículos de emergencia en tiempo real. Para obtener la huella digital de una sirena Hi-Lo se aplicaron diversas técnicas de procesamiento digital de señales y simbolización de señales, mismas que fueron contrastadas contra un clasificador de audio basado en una red neuronal profunda, partiendo de un mismo conjunto de 280 audios de sonidos ambientales y 38 audios de sirenas Hi-Lo. En ambos métodos se evaluó su precisión a partir de una matriz de confusión y diversas métricas. La precisión del algoritmo DSP desarrollado presentó una mayor capacidad para discriminar entre la señal y el ruido, en comparación con el modelo CNN.

**Palabras Clave:** detección de vehículos de emergencia, huella digital de audio, simbolización de series temporales, detección de eventos acústicos, espectrogramas de Mel.

**ABSTRACT.**
In emergency situations, the movement of vehicles through city streets can be problematic due to vehicular traffic. This paper presents a method for detecting emergency vehicle sirens in real time. To derive a siren Hi-Lo audio fingerprint it was necessary to apply digital signal processing techniques and signal symbolization, contrasting against a deep neural network audio classifier feeding 280 environmental sounds and 38 Hi-Lo sirens. In both methods, their precision was evaluated based on a confusion matrix and various metrics. The precision of the developed DSP algorithm presented a greater ability to discriminate between signal and noise, compared to the CNN model.

## 1. INTRODUCCIÓN.

Existen diversos tipos de sirena en los vehículos de emergencia (VE). En situaciones de emergencia, el desplazamiento de un VE por las vialidades de una urbe puede ser problemático debido al tráfico vehicular. Es entonces que un sistema de detección de sirenas de vehículos de emergencia (EVD) puede asistir, por ejemplo, en los sistemas de asistencia en el manejo [1]. Los sonidos de sirena EV se pueden separar en tres categorías: ambulancias, camiones de bomberos y patrullas de policía. Cada país tiene su regulación para los diversos efectos de sonido de una sirena y respectivo rango de frecuencias [1]. La Organización Internacional de Normalización (ISO) establece una norma para sonidos de emergencia de alertas de evacuación, desastres naturales y sirenas. Dicha norma estipula el rango de frecuencias para sonidos de sirenas, que abarca un rango de 500 Hz a 2,500 Hz, recomendando la inclusión de dos tonos dominantes dentro del rango de 500 Hz y 1500 Hz [2]. Mientras que, en México, la norma NOM-034-SSA3-2013 solo menciona el rango de los niveles en decibeles promedio para la sirena de una ambulancia, entre 120 y 130 decibeles.

Los tres efectos más comunes para un sonido de sirena generado electrónicamente son: Hi-Lo *(two-tones),* yelp *y* wail [3-7]. Para la identificación EVD mediante una huella digital de audio (*audio fingerprint*) es necesario aplicar una serie de herramientas y técnicas como lo son: el procesamiento digital de señales (DSP) [4, 8-11], y clasificadores de audio basados en modelos de redes neuronales profundas (*deep learning neural networks,* DNN) [1, 6, 12], así como la representación del audio en el espectrograma de Mel, el cual es ampliamente utilizado en las tareas de clasificación de sonidos ambientales [1, 6, 11-12].

El presente trabajo presenta un novedoso método EVD en tiempo real basado en la simbolización de señales unidimensionales en series temporales para su posible aplicación en vehículos automotores como un sistema avanzado de asistencia al conductor (*advanced driver assistance system*, ADAS) que auxilie en una conducción más cooperativa orientada a agilizar la circulación urbana de vehículo EV [13].

## 2. TRABAJOS RELACIONADOS.

Entre los trabajos relacionados destaca [4], donde se implementa un algoritmo detector de vehículos de emergencia (EV) con ayuda de DSP por medio de la identificación de sus tonos dominantes (*pitch detection*), su método consta de dos etapas: la técnica MDF (*module difference function*) para clasificar cada porción del audio en sonidos con tonos definidos y la segunda etapa consiste en buscar los picos de cada tono para detectar patrones en el tiempo similares a una sirena Hi-Lo. En [1] presentan tres modelos de redes neuronales diferentes: (*fully connected model*, FCN), convolucional (CNN) y recurrente (RNN). Dichos modelos ML son capaces de extraer los rasgos característicos de los EVs mediante los coeficientes cepstrales de frecuencias de Mel (MFCC). En [7] se utilizó un análisis espectral mediante la transformada rápida de Fourier (FFT) para identificar sirenas y tratar de estimar la distancia a la que se encuentra el vehículo EV. Se menciona como limitante que el sonido del motor del vehículo genera problemas para realizar la detección cuando la potencia del sonido es mayor al sonido de la sirena. En [12] se realizó una comparativa entre los modelos CNN AlexNet y GoogleNet utilizando tres diferentes representaciones para los audios de entrada: espectrograma de frecuencias, MFCC y gráficos de recurrencia cruzada (CRP).

En análisis de series de tiempo, la correlación cruzada de dos secuencias de señales puede ser aplicada para encontrar su similaridad midiendo su desplazamiento relativo, sin embargo, en la práctica presenta varias limitaciones [14, 15]. El análisis simbólico de series en el tiempo para identificar rasgos de interés fue aplicado en [16] usando autómatas de estado finito probabilísticos (PFSA) y mediante la técnica SAX [17], que se destaca por mapear segmentos de señal como símbolos en una cadena de texto en base a los promedios de la distribución normal de cada segmento.

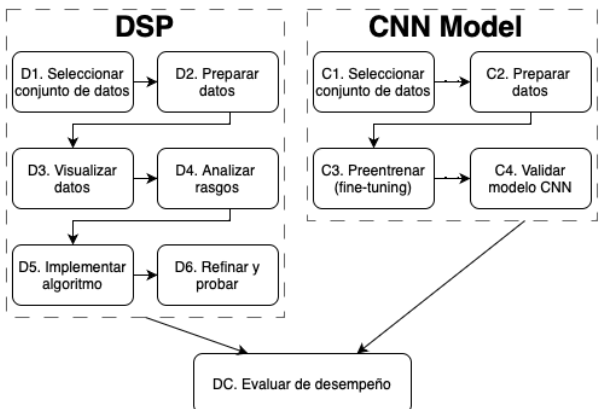

Figura 1. Metodología para ambos métodos: DSP y modelo CNN.

3. **METODOLOGÍA.**
El método propuesto para la detección de sirenas Hi-Lo fué comparado contra un modelo de red neuronal convolucional (CNN), siguiendo el procedimiento como se detalla en la Fig. 1. En ambos modelos parten del mismo conjunto de datos, pero tienen procesos específicos para la selección y preparación de los mismos. El método propuesto se caracteriza por utilizar transformaciones propias de DSP, aportando una simbolización de la señal para la extracción y detección de rasgos usando expresiones regulares que se detallan en las secciones correspondientes. Al final se efectuó una comparativa de la evaluación del desempeño obtenido en cada uno de los métodos de detección y se indican las conclusiones obtenidas.
Para desarrollar el algoritmo de DSP y el entrenamiento del modelo CNN se siguieron las etapas de la Fig. 1. Selección del conjunto de datos y preparación de los clips de audio. Para el algoritmo de DSP fue necesaria la selección del método de extracción de rasgos característicos para el entrenamiento del modelo, así como para la representación de la firma acústica del audio. Proponer y afinar el algoritmo detector de sirenas. Para el entrenamiento del modelo pre-entrenado fue necesario unas conversiones en el formato del audio digital. Finalmente se procedió a realizar las pruebas de desempeño y validación de los resultados obtenidos.

4. **DESARROLLO.**
A continuación, se explica de forma breve cada una de las etapas involucradas en el desarrollo de ambos modelos de detección de sirenas Hi-Lo de forma separada, donde el método propuesto basado en DSP abarca las secciones D1-D6 y el método basado en un modelo CNN incluye los pasos C1-C4. Al final se detalla el análisis comparativo del rendimiento. A continuación, se describen las etapas de desarrollo del algoritmo de detección basado en DSP y expresiones regulares:

**D1. Selección de los conjuntos de datos:** se eligieron dos conjuntos de datos reconocidos para la clasificación de sonidos ambientales:
- **ESC-50:** consiste en 2,000 clips de audio que corresponden a 50 clases diferentes, cada uno con una duración de 5 segundos. El conjunto de datos es muy uniforme en todos sus clips con una tasa de muestreo de 44.1 kHz [18].
- **UrbanSound8k:** consiste en 8,732 clips divididos en 10 clases de diferentes sonidos urbanos. Sus audios tienen una duración de 4 seg. o inferior, y una tasa de muestreo varía de 11 kHz a 96 kHz, y un *bitrate* de 8 a 32 bits [19].

Ambos conjuntos de datos tienen una validación cruzada organizados por carpetas de cada clase [20-24].

**D2. Preparación de datos**: En esta etapa se realizó la organización y selección de los audio-clips de interés, siendo la clase de sirena nuestro objetivo principal. Se obtuvieron alrededor de 40 ejemplos de sirenas Hi-Lo entre ambos conjuntos de datos para sirenas. Se descartaron varios clips por sus variaciones en la tasa de muestreo, así como en duraciones inferiores a dos segundos.

**D3. Representación de los datos:** La extracción y análisis de los rasgos característicos de las señales de audio se efectuaron dentro del dominio de la frecuencia, obteniendo el espectrograma logarítmico de Mel (*Log-Mel spectrogram*) mediante la librería de Python para procesamiento de audio y música Librosa [25], con los siguientes parámetros: filtro de Hann con 1,024 muestras por FFT y un tamaño de ventana de 320 muestras. Dicha transformación expresa la intensidad del audio en forma logarítmica (dB) para el rango de frecuencia de interés en función del tiempo [26].

**D4. Análisis de rasgos DSP:** Una sirena Hi-Lo se compone de dos tonos que oscilan entre ellos de manera cíclica, siendo uno de los tonos de mayor frecuencia y otro de menor frecuencia. La figura 2.2 nos muestra la representación visual del espectrograma Log-Mel para un audio de una sirena Hi-Lo.

**D5. Algoritmo DSP:** En este apartado se presentan los principales rasgos de la señal considerados para su identificación y el algoritmo propuesto para la detección de sirenas Hi-Lo incluyendo algunos ejemplos para elementos clave del algoritmo propuesto.

Entre las principales características de interés consideradas para una señal acústica de sirena Hi-Lo se encuentran los siguientes:

$R_1$: banda de frecuencias en intervalo $f_a$-$f_b$
$R_2$: umbral frecuencia dominante ($dB_{min}$)
$R_3$: presencia de tonos Hi-Lo.
$R_4$: brecha tonal *a-b* (*gap*)
$R_5$: variabilidad nivel tono dominante $dB_{min}$
$R_6$: discontinuidad máxima para tonos Hi-Lo
$R_7$: periodicidad mínima esperada
$R_8$: patrón esperado de regularidad periódica

Enseguida se describe el pseudo-código del algoritmo propuesto para identificar sirenas Hi-Lo mediante la secuenciación simbólica de series de tiempo de señales unidimensionales para la extracción e identificación de rasgos usando expresiones regulares, teniendo como parámetro de entrada una señal de audio, obteniendo como resultado el *status* de identificación de señal Hi-Lo como un valor lógico (*false, true*):

1. Obtener espectrograma Log-Mel del audio.
2. Filtrar banda $f_a$-$f_b$ de interés según criterio $R_1$.
3. *Encoder*$_1$: simbolizar cada segmento de forma binaria en base a $R_2$.
4. Generar vector *v* con frecuencias dominantes en cada segmento.
5. Generar histograma de las frecuencias dominantes presentes en *v*.
6. Si no existen tonos Hi-Lo (criterio $R_3$) salir reportando valor *false*.
7. Ciclo de detección (máx. 4 ciclos):
   1. Seleccionar tonos Hi-Lo con mayor conteo.
   2. Satisfacer criterio $R_4$; si falla, eliminar tono más frecuente y ciclar
   3. *Encoder*$_2$: codificar respetando criterio $R_5$
   4. Reconstruir señal mediante expresión $C_4$ que incorpora $R_6$; dato de entrada sin cambios si falla la reconstrucción.
   5. Identificar patrón $C_5$ bajo criterio $R_7$; si falla, eliminar mayor tono (armónico) y ciclar.
   6. *Encoder*$_3$: codificar según expresión $C_6$ y verificar criterio $R_8$ usando $C_7$; si falla, ciclar. Si se cumple terminar y salir con valor *true*.
8. Si concluye ciclo, regresar valor *false*.

A continuación, se explican las principales transformaciones de la señal utilizando ejemplos. Primero, se obtiene el espectrograma Log-Mel del audio (Fig. 2) indicando la intensidad del audio (dB) entre el intervalo de frecuencias 20 Hz a 2,560 Hz (eje Y en escala $Log_2$) para cada segmento de tiempo (*frames* del eje X) en matriz Log-Mel de [64×65]. Dadas las frecuencias límite de los tonos Hi-Lo $f_a$ y $f_b$ se extrae la franja $f_a$-$f_b$ de interés mediante la expresión:

$$cR_1 = \{ x \in LogMel \mid f_a \leqslant x \leqslant f_b \} \quad (C_1)$$

La primera simbolización de señal (*encoder*$_1$) transforma a dígitos binarios (0,1) los niveles de intensidad de la franja $f_a$-$f_b$ mediante:

$$cEnc_1 = \begin{cases} 1 & \text{if } |x| > dB_{min} \text{ where } x \in cR_1 \\ 0 & \text{otherwise} \end{cases} \quad (C_2)$$

Para la segunda simbolización (*encoder*$_2$) se considera el siguiente alfabeto:

$$\sum = \{a,b,-\} \quad (Alphabet)$$

Donde el símbolo terminal a representa un tono Hi, b un tono Lo y el símbolo – engloba valores fuera del rango de interés, codificado mediante:

$$cEnc_2 = \begin{cases} a & \text{if } x \in R_5^{hi} \\ b & \text{if } x \in R_5^{lo} \\ - & \text{otherwise} \end{cases} \quad (C_3)$$

Luego de encontrar histograma con frecuencias dominantes ($R_2$), identificar candidatos para tonos Hi-Lo ($R_3$) y verificar si dichos tonos cumplen $R_4$, obtenemos una secuencia de tonos tal que, como ejemplo, una secuencia `aabbaabb` puede indicar la presencia de una sirena Hi-Lo que consta de dos ciclos. Sin embargo, dado que pueden existir errores de muestreo (ruido, latencia, conversión ADC), en ocasiones puede ser necesario reconstruir la señal considerando las siguientes discontinuidades ($R_6$):

$$cR_6 = \text{a-a|a--a|b-b|b--b} \quad (C_4)$$

Al reconstruir la señal en el paso 7.4, por ejemplo, para una sección `a--a` obtenemos el período completo `aaaa`, siendo así más probable detectar sirenas Hi-Lo verificando una periodicidad mínima ($R_7$) mediante:

$$cT_{min} = \text{(a+b+)+} \quad (C_5)$$

Sin embargo, dicha expresión no garantiza cumplir el criterio de regularidad periódica $R_8$, es decir, si ambos tonos se ajustan a un determinado rango de variación periódica, ya que, en la práctica, pueden existir ligeras variaciones por ruido, precisión, temperatura o recortes por inicio o fin del muestreo de señal. Para efectuar dicho análisis es necesario simbolizar de nuevo la señal (*encoder*$_3$) bajo el siguiente formato:

$$cEnc_3 = \text{(a|b|-)d+} \quad (C_6)$$

Por ejemplo, para la cadena `aaabbaabbaab`, se obtiene la cadena `a3b2a2b2a2b1`. Luego se clasifican y ordenan dichos tonos, es decir, [a3,a2,a2] y [b2,b2,b1] y se procede a eliminar de manera recursiva las ocurrencias atípicas hasta una dimensión *n*=2, i.e. [a2, a2] y [b2,b2]. Luego se calcula la media x̄ y varianza $\sigma^2$ de cada lista y se considera como criterio de aceptación de regularidad periódica $R_8$, la siguiente condición:

$$cT(x) = \overline{x} > \sigma^2(x) \quad (C_7)$$

Finalmente, al cumplir todos los criterios ($R_{1-8}$) mediante las expresiones $C_{1-7}$, el algoritmo considera exitosa la identificación de una sirena Hi-Lo.

**D6. Pruebas DSP:** las constantes definidas para realizar exitosamente las pruebas de validación del algoritmo tenemos:

$R_1$: $f_a$ = 1500 Hz, $f_b$ = 700 Hz
$R_2$: $dB_{min}$ = 20 dB
$R_4$: $gap$ = 122 Hz
$R_5$: ±31 Hz
$R_6$: 2 ciclos @ dur ≈ 4 segs.
$R_7$: 1-3 *frames*

**C1. Selección de conjunto de datos para modelo CNN:** Para el modelo CNN pre-entrenado se trabajó principalmente con ESC-50 para la selección de 7 clases para re-entrenar el modelo CNN (*transfer learning*), constando de 40 audios para cada clase tales como: sirenas, animales, claxon de carro, avión, lluvia, viento. Con UrbanSound8k se obtuvieron 280 audios de clases independientes a sirenas, como: ladridos de perro, niños jugando, aire acondicionado, claxon de carro, motor en marcha, sonidos de construcción (taladros y martillos). Se recopilaron 38 clips de sirenas Hi-Lo, los cuales no estuvieron presente en el entrenamiento del modelo, posteriormente se utilizaron para las pruebas comparativas entre los dos métodos.

**C2. Preparación de datos:** También se seleccionaron de manera aleatoria 280 audios de entre el resto de las clases para nuestra categoría de audios de no-sirenas. Esta parte fue importante para el desarrollo del entrenamiento del modelo CNN, así como la validación del algoritmo DSP y modelo CNN. Los clips de audio se encuentran en su gran mayoría de conjunto de datos en formato digital de audio WAVE [27], el cual no es compatible con la herramienta CreateML de Xcode para entrenar el modelo CNN, por lo tanto, fue necesaria una conversión del audio a formato CAF [28] con ayuda de la herramienta `afconvert` que está presente en el sistema operativo de MacOS [29].

**C3. Entrenamiento (*fine-tuning*):** Al ser un modelo pre-entrenado CNN [30], entrenamos con las clases de interés, este entrenamiento llevó los siguientes parámetros: Extractor de rasgos: *Audio Feature Print*, iteraciones: 105, duración del ventaneo: 0.975s, *overlap*: 50%.

**C4. Pruebas modelo CNN:** CreateML incluye un apartado de evaluación del modelo entrenado, para lo cual se utilizaron 4 clips de cada una de las clases para realizar la validación interna del modelo, obteniendo 99.8% de precisión de entrenamiento con 105 iteraciones, y 96.4% de precisión en la validación.

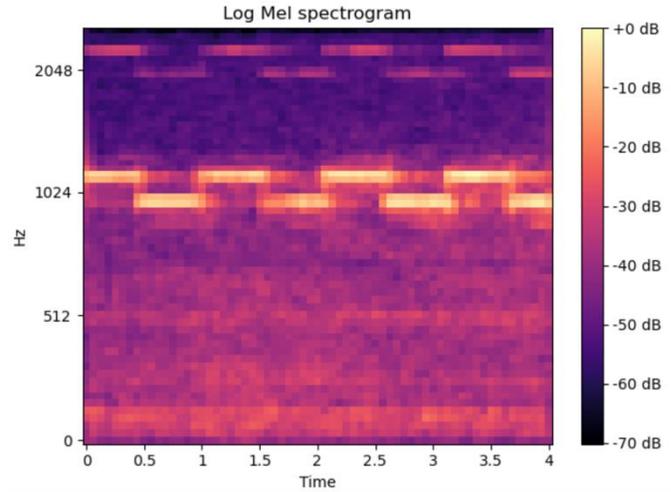

Figura 2. Espectrograma logarítmico de Mel para sirena con efecto Hi-Lo.

**DC. Evaluación de desempeño de ambos modelos:** Para los 280 audios con clases diferentes a sirenas, se contabilizaron los audios que fueron clasificados correctamente como audios no-sirena (TN), así como los que fueron clasificados incorrectamente como audios de sirenas (FP), de la misma forma, de los 38 clips para sirenas Hi-Lo, se contabilizaron los audios correctamente clasificados como sirenas (TP), así como los clasificados incorrectamente como audios de no-sirena (FN) tanto para el modelo de CNN como para el algoritmo DSP.

Posteriormente, con el propósito de evaluar el rendimiento de ambos modelos de clasificación, a partir de dicha información se construyó la matriz de confusión de cada modelo y se calcularon las métricas de desempeño precisión, sensibilidad (*recall*), especificidad y puntuación F1. Además, con base en la teoría de detección de señales (SDT), se calculó la capacidad de discriminación del sistema d' y el criterio de decisión de la siguiente forma:

$$Precisión = TP/(TP + FP)$$

$$Sensibilidad\ (recall) = TP/(TP + FN)$$

$$Especificidad = TN/(TN + FP)$$

$$F1 = \frac{2(Precisión\ x\ Sensibilidad)}{(Precisión + Sensibilidad)}$$

$$d' = Z(proporción\ de\ TP) - Z(proporción\ de\ FP)$$

$$Criterio = \frac{Z(proporción\ de\ FP) + Z(proporción\ de\ FP)}{-2}$$

## 5. Resultados.

La matriz de confusión para el modelo CNN, así como para el algoritmo DSP se muestran en las tablas 1 y 2 respectivamente.

**Tabla 1**. Matriz de confusión del modelo CNN.

|  |  | Detección Sirena Hi-Lo | |
|---|---|---|---|
|  |  | Negativo | Positivo |
| 38 sirenas Hi-Lo | Positivo | 11 (FN) | 27 (TP) |
| 280 no-sirenas | Negativo | 266 (TN) | 14 (FP) |

**Tabla 2**. Matriz de confusión del algoritmo DSP.

|  |  | Detección Sirena Hi-Lo | |
|---|---|---|---|
|  |  | Negativo | Positivo |
| 38 sirenas Hi-Lo | Positivo | 2 (FN) | 36 (TP) |
| 280 no-sirenas | Negativo | 273 (TN) | 7 (FP) |

El modelo CNN no detectó 11 sirenas (FN) y detectó como sirenas 14 audios sin sirenas (FP). Por otro lado, el modelo DSP no detectó 2 sirenas (FN) e identificó incorrectamente 7 audios como sirenas dentro del conjunto de datos de no-sirenas (FP).

En la tabla 3, se resumen las métricas evaluadas, en donde se observa que el modelo DSP presentó una menor tasa de error en comparación con el modelo de CNN (0.03 vs 0.08). De igual manera, el modelo de DSP registró una mayor sensibilidad o tasa de verdaderos positivos, detectando el 95% de los audios de sirena, en comparación con el modelo CNN que detectó el 71%.

**Tabla 3**. Métricas de desempeño para modelo CNN y DSP.

| Métrica | Modelo CNN | DSP |
|---|---|---|
| Tasa de Error | 0.08 | 0.03 |
| Sensibilidad | 0.71 | 0.95 |
| Especificidad | 0.95 | 0.98 |
| Precisión | 0.66 | 0.84 |
| Puntuación F1 | 0.68 | 0.89 |
| Discriminación d' | 2.20 | 3.58 |
| Criterio de decisión | 0.55 | 0.17 |

En cuanto a la especificidad, que se refiere a la proporción de audios de no-sirenas que el modelo clasificó correctamente como no-sirena, de entre todas las predicciones negativas que hizo, el modelo DSP resultó más alto con un 98% en comparación con un 95% del modelo CNN, el modelo CNN detectó erróneamente 14 sirenas que en realidad no lo eran, el doble del modelo DSP que presentó 7 falsos positivos. De la misma forma, la precisión, que se refiere a la proporción de audios de sirenas Hi-Lo que el modelo clasificó correctamente como audio de sirena, de entre todas las predicciones positivas que hizo resulto mayor en el modelo DSP (0.84) en comparación con el modelo CNN (0.66), el modelo DSP presentó un mayor equilibrio entre la sensibilidad y la precisión (puntuación F1 = 0.89 vs 0.68).

En lo referente a la capacidad para discriminar, el modelo DSP presentó un valor mayor de d' (3.58 vs 2.20) en comparación con el modelo CNN, lo que significa una mayor capacidad para discriminar entre la señal y el ruido. El valor del criterio fue menor para el modelo DSP, es decir presentó un criterio más liberal, lo que significa que es más propenso a identificar un audio como sirena Hi-Lo (TP), esto resulta ser útil para detectar señales débiles pero que son importantes, sin embargo, esta mayor sensibilidad se presenta acompañada de un aumento en las falsas alarmas, pero esto resulta ser menos importante.

**Observaciones**. El método de detección propuesto es una alternativa de la técnica SAX (*Symbolic Aggregate Approximation*) [17], introducida en 2003 [31], donde afirman presentar una novedosa representación simbólica de alto nivel de señales en series de tiempo, diferente a las técnicas basadas en transformadas de Fourier, *wavelets, eigen waves*, modelos polinomiales (*piecewise polynomial models*). Gracias a la representación adaptativa de rasgos de señal como patrones de texto, es posible aplicar algoritmos de procesamiento de texto, en específico para el método aquí descrito, el uso de expresiones regulares a partir de espectrogramas de Mel para detectar patrones en señales codificadas como cadenas de texto, a diferencia de otros métodos basados en ciencia de datos, tales como: *clustering, classification, indexing, summarization, trees, anomaly detection* [17].

**Trabajo a futuro.** La reducida cantidad y calidad de las muestras de prueba puede afectar las pruebas de precisión; al revisar la calidad de las muestras, se encontró que las señales no detectadas no satisfacen alguno de los criterios $R_1$-$R_8$, por lo que a futuro se recomienda incorporar una mayor cantidad de sonidos de sirenas Hi-Lo de mayor calidad, de preferencia de vehículos locales, respetando dichos criterios y siguiendo un estricto protocolo de grabación de audio. Durante las pruebas del algoritmo DSP se llevaron a cabo diversas iteraciones para ajustar los valores de los criterios $R_1$-$R_8$, mismos que brindan un modelo más moldeable y auto-explicable en comparación con los modelos CNN. En la práctica, esto ofrece la posibilidad de ajustar dichos parámetros para mejorar sus capacidades de detección EV de una urbe específica, por lo que se recomienda realizar dichas lecturas, experimentos y ajustes en un futuro inmediato [32]. Otro importante desarrollo como trabajo a futuro, consiste en entrenar una red CNN a partir de los espectrogramas de Log-Mel para aplicarlo a la detección de radioisótopos. Además, queda pendiente efectuar las mediciones del rendimiento del algoritmo DSP tanto para una aplicación móvil (ARM 64-bits), como en una tarjeta de desarrollo basado en un microcontrolador de bajo costo (ARM 32-bits), para así demostrar su viabilidad y rango de aplicaciones, e.g. controlador ADAS [32].

## 6. Conclusiones.

Se reporta un algoritmo desarrollado para detectar sirenas Hi-Lo para vehículos de emergencia. El modelo es auto-explicable y está basado en espectrogramas de Mel y expresiones regulares. Se obtuvo una precisión del 84% para la implementación del modelo DSP propuesto, ofreciendo una tasa inferior de falsos positivos en comparación con el modelo CNN de prueba, gracias en buena medida a la facilidad de ajustar dicho modelo por medio de criterios de decisión comprensibles y bien definidos ($R_1$-$R_8$ y $C_1$-$C_7$), destacando también su capacidad para discriminar entre la señal de interés y el ruido presentes en los audios de prueba.